# Modelling mechanical percolation in graphene-reinforced elastomer nanocomposites

Mufeng Liu, Ian A. Kinloch, Robert J. Young, Dimitrios G. Papageorgiou*

*School of Materials and National Graphene Institute, University of Manchester, Oxford Road, M13 9PL, Manchester, United Kingdom*

*Corresponding author: dimitrios.papageorgiou@manchester.ac.uk (Dimitrios G. Papageorgiou); Tel. (+44) 161 306 1478.

**Abstract**

Graphene is considered an ideal filler for the production of multifunctional nanocomposites; as a result, considerable efforts have been focused on the evaluation and modeling of its reinforcement characteristics. In this work, we modelled successfully the mechanical percolation phenomenon, observed on a thermoplastic elastomer (TPE) reinforced by graphene nanoplatelets (GNPs), by designing a new set of equations for filler contents below and above the percolation threshold volume fraction ($V_p$). The proposed micromechanical model is based on a combination of the well-established shear-lag theory and the rule-of-mixtures and was introduced to analyse the different stages and mechanisms of mechanical reinforcement. It was found that when the GNPs content is below $V_p$, reinforcement originates from the inherent ability of individual GNPs flakes to transfer stress efficiently. Furthermore, at higher filler contents and above $V_p$, the nanocomposite materials displayed accelerated stiffening due to the reduction of the distance between adjacent flakes. The model derived herein, was consistent with the experimental data and the reasons why the superlative properties of graphene cannot be fully utilized in this type of composites, were discussed in depth.

1. Introduction

Thermoplastic elastomers (TPEs) that utilize physical crosslinks to achieve elastomeric characteristics, are easier to process than conventional rubbers but are limited by their poor mechanical properties relative to thermoplastics [1]. As a result, TPEs are commonly reinforced by inorganic fillers [2-5]. Since the isolation of single layer graphene [6] with its exceptional mechanical properties (Young's modulus ~1 TPa, strength ~130 GPa) [7, 8], several studies have focused on the mechanical enhancement of elastomer-based nanocomposites, reinforced by graphene [3, 9-18].

A number of micromechanical theories have been used in the literature to describe the reinforcement of polymers by 2D materials [18]. Classical theories such as the Guth-Gold theory [2] based on hydrodynamics and more modern ones, such as the jamming theory [3, 19] based on the percolation phenomenon, have been utilized to explain the stiffening mechanisms in elastomer/graphene composite systems [4, 20]. The corresponding theoretical analysis has shown good consistency with the experimental results [3, 4, 11, 12, 15, 20]; nevertheless, some questions still stand when moving from the microscopic to the macroscopic scales in order to explain fully the mechanisms of mechanical enhancement. In our recent work, we established the mechanisms of reinforcement of polymers by graphene nanoplatelets and showed that for elastomers possessing low shear modulus, the reinforcing efficiency of graphene nanoplatelets is dependent upon the aspect ratio and the volume fraction of the filler, whilst virtually independent of the filler modulus ($E_f$) [21].

In the present study, a new analytical method is developed by the combining shear-lag and the rule-of-mixtures theories, along with the mechanical percolation phenomenon, in an attempt to evaluate the stiffening mechanism in TPE/GNP nanocomposites. A semicrystalline polyether block amide thermoplastic elastomer was employed to prepare nanocomposites with two types of GNPs of different diameters, by melt mixing in an internal mixer. The degree of crystallinity of the matrix and the composites was obtained by differential scanning calorimetry (DSC) and X-ray diffraction (XRD). The microstructure of the nanocomposites was characterized by scanning electron microscopy (SEM), while the mechanical properties were investigated by tensile testing and *in situ* deformation under a Raman spectrometer. A detailed theoretical analysis was carried out to determine the different mechanisms and stages of mechanical reinforcement of the produced TPE/GNP nanocomposites.

## 2. Experimental methods

*2.1 Materials and preparation*

A commercially-available thermoplastic elastomer (Pebax® 7033) was purchased from Arkema, Inc. and used as received. Pebax® is a plasticizer-free polyether block amide. Graphene nanoplatelets (GNPs) with nominal lateral diameters of 5 and 25 μm (M5 and M25) (according to the supplier) and average thicknesses in the range of 6–8 nm were purchased from XG Sciences, Inc. Lansing, Michigan, USA and used as received. Composites with 1, 5, 10, and 20% by weight of the GNPs were prepared by melt mixing in a Thermo Fisher HAAKE Rheomix internal mixer at 220 °C and 50 rpm for 10 minutes. The lumps of composites were afterwards injection moulded in a HAAKE Minijet Piston Injection Moulding System in order to prepare dog-bone shaped specimen. Based on the fact that the nanocomposites containing the M25 GNPs displayed higher values of Young's modulus (compared to M5 GNPs), a wide range of composites with various M25 GNP loadings (1, 2.5, 3.3, 5, 6.7, 7.5, 10, 11.1, 12.5, 13.3, 15, 16.6, 17.5, 18.9, 20 wt%) were additionally prepared in order to obtain sufficient data for theoretical analysis and modelling of the mechanical percolation phenomenon. The samples throughout the paper will be coded based on the matrix (TPE), the diameter and the weight content of the filler. For example, TPE-M5-GNP1 refers to the TPE matrix reinforced by 1 wt% of GNPs, whose average diameter is 5 µm.

*2.2 Characterization of the nanocomposites*

The final filler contents of GNPs in the nanocomposites were evaluated by thermogravimetric analysis (TGA) using a TA Q500 TGA instrument. The samples were heated from 25 °C to 800 °C under a 50 mL/min flow of $N_2$ at 10 °C/min. Three samples were tested for each material in order to ensure reproducibility of the results.

The XRD diffractograms were obtained using a PANalytical X'Pert3 diffractometer with Cu Kα radiation operated at 40 kV and 40 mA. The 2-theta angle range was selected from 5° to 90° with a step size of 0.03° and a step time of 180 s.

A TA Instruments Q100 DSC was used to investigate the melting and crystallization behaviour. Samples of about 10 mg were heated, cooled and re-heated between -90 and 200 °C using a heating/cooling rate of 10 °C/min, under a nitrogen flow of 50 ml/min.

The morphology of the neat polymer and the microstructure of the nanocomposites were examined using SEM. The images of the cryo-fractured samples were acquired using a high-resolution Philips XL30 Field Emission Gun Scanning Electron Microscope (FEGSEM) operated at 6 kV.

Stress–strain curves were obtained using dumbbell-shaped specimen in an Instron 3365 machine, under a tensile rate of 10 mm·$min^{-1}$ with a load cell of 5 kN, in accordance with the ASTM D638 standard. An extensometer with a gauge length of 20 mm was used to measure the strain precisely.

Raman spectra were obtained using a Renishaw InVia Raman spectrometer with a laser wavelength of 633 nm and a x50 objective lens, which produces a laser spot with a diameter in the order of 1–2 µm. The Raman 2D band shift of the injection moulded samples was studied following the application of strain on the nanocomposites. The strain was applied with a four-point bending rig from 0 to ~ 2.7% and determined by a resistance strain gauge attached to the surface of samples. Spectra were taken every ~0.1 % strain. The Raman spot was focused on the same point of a single flake on each sample surface. All spectra were fitted with a single Lorentzian curve.

## 3. Results

*3.1 Volume fraction and degree of crystallinity*

The volume fraction of the fillers within the composites was assessed using TGA. The mass residue along with the volume fractions of the filler were calculated by equation S1 and are given in the Table S1 (Supporting Information). The final mass fractions were very close to the nominal mass fractions of the nanocomposites.

The degree of crystallinity of the neat polymer and the nanocomposites was characterised using both XRD and DSC (Figure S1). In the X-ray diffractograms, the matrix displays only one characteristic peak which is present at 2θ $\approx$ 21.4°, showing a reflection from the (001) plane, corresponding to the $\gamma$ phase of polyamide 12 (PA12) [22]. The XRD patterns of GNPs display a sharp and strong peak at 2θ $\approx$ 26° consistent with reflections from the (002) plane of graphite. The degree of crystallinity can be calculated by $X_c=A_c/(A_c+A_\alpha)$ for XRD, where $A_c$ and $A_\alpha$ are the areas under the crystalline peaks and amorphous halo, respectively. The results can be seen in Table S2. In addition, the DSC results suggest that the melting point ($T_m$) of all samples is at around 160 °C and the presence of GNPs did not affect the $T_m$. The degree of crystallinity from DSC was calculated from the ratio $X_c= \Delta H_f/\Delta H_f^0 \times 100\%$, where $\Delta H_f$ is enthalpy of fusion of the sample and $\Delta H_f^0$ is the enthalpy of fusion of 100% crystalline PA12 in Pebax (65 J/g [23]). As can be understood from the application of both techniques, the presence of GNPs did not alter the crystallinity of the matrix significantly.

*3.2 Microstructure of the nanocomposites*

The SEM images of cryo-fractured cross-sections of the nanocomposite samples reinforced with M5 GNPs, at different filler fractions, can be seen in Figure 1. The corresponding images of M25-GNP-reinforced nanocomposites are shown in Figure S2. It can be seen that a homogeneous dispersion of the flakes was achieved as a result of the mixing procedure. Moreover, the injection moulding procedure attributed a preferred orientation of the flakes in the axial direction of the samples, due to the fountain flow mechanism [14]. Overall, the distance between individual flakes in the vertical direction (the direction of the surface normal of the flakes) in the images, is reduced with increasing filler loading. A measurement of the distances between neighbouring flakes was carried out based on more than 100 flakes for each sample and shown in Figure S3. This reduction in inter-flake distance can activate a pronouncedly enhanced mechanical performance, which will be discussed in detail later. It should be pointed out that both batches of GNPs seem to include a number of smaller flakes, which decrease significantly the average lateral size quoted by the manufacturer, as was also identified in a previous work from our group [21].

The high magnification SEM images in Figure S4 reveal a good interface between the flakes and the matrix. However, the morphologies of the flakes in the bulk nanocomposites also indicate stacking and the formation of agglomerates among the flakes at higher filler contents (TPE-GNP20 sample in Figure 1d and Figure S2), that are known to reduce the reinforcing efficiency by subsequently reducing the average aspect ratio of the flakes. In addition, some flakes can be seen forming looped or folded morphologies, similar to the ones observed in our previous study for a different thermoplastic elastomer matrix [14]. The pre-existing folds of the GNPs along with the high shear that develops during the mixing process, can be both attributed for the folding and the bending of the nanoplatelets within the nanocomposites.

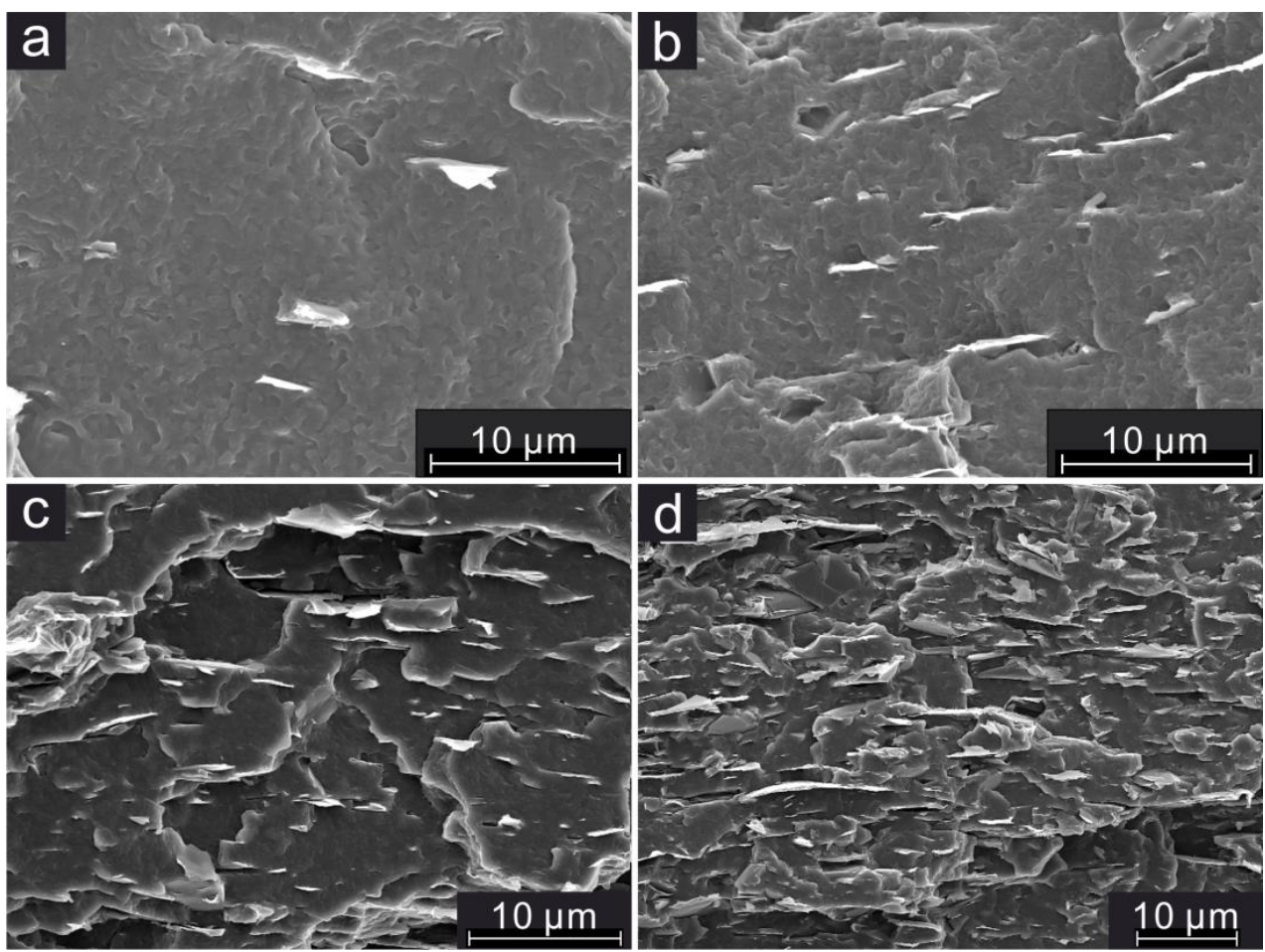

**Figure 1.** SEM images of the composites (a) TPE-M5-GNP1, (b) TPE-M5-GNP5, (c) TPE-M5-GNP10, (d) TPE-M5-GNP20.

*3.3 Tensile testing*

The mechanical properties of the produced nanocomposites were initially evaluated by tensile testing. Typical stress-strain curves of all nanocomposite samples are shown in Figure S5. It is interesting to observe that the addition of GNPs into the matrix alters the tensile behaviour of the original TPE. More specifically, the yield point shifts to higher stress and lower strain with the increase of the filler loading, indicating that the mobility of the macromolecular chains is restricted [3]. Moreover, the samples with high filler loadings (above 5 wt%) tend to display only elastic deformation, while finally at the highest GNP content (20 wt%), the stress-strain curves of the semi-crystalline TPE nanocomposites are characteristic of a brittle polymer. From the mechanical properties results presented in Figure 2a and b, it can be seen that both the yield strength and the tensile modulus increase with increasing filler content, indicating good stress transfer efficiency through shear at the filler-matrix interface. Overall, the M5 GNPs give rise to higher improvements in the yield strength, as a result of their better dispersion and consequently filler-matrix adhesion, while M25 GNPs contribute slightly more to the stiffness of the materials indicating better stress transfer. However, the addition of both types of GNPs results in reductions of the elongation at break (apart from the samples filled with only 1 wt% of GNP). This is commonly observed in polymer nanocomposites where agglomerates of fillers can initiate failure during elongation [18].

The results of the Young's modulus values against the filler loading clearly display a superlinear increase rather than a linear one, implying that the higher loadings of GNPs exert an additional enhancement in the modulus, compared to the lower loadings. This is consistent with previous findings for different types of elastomer matrices reinforced by 2D materials [3, 4, 20]. Based on this finding, we prepared a number of TPE-M25-GNP nanocomposites with various filler loadings, in order to establish the transitional turning point in the modulus versus loading graph. It can be clearly seen in Figure 2c that the slope of the linear fit of the normalized modulus of the composite against the volume fraction of the filler is significantly higher for loadings above ~ 5 vol%. Quantitatively, the slope of the data below 5 vol% is 24, whereas the corresponding slope for the data above 5 vol% is 61. This phenomenon observed for elastomer composites, was defined as accelerated stiffening by Guth and Gold [2], where the rate of the increase of the composite modulus increases with increasing filler volume fractions. Accelerated stiffening has been observed in elastomer composites for a variety of fillers including carbon black [2], clays [3, 4] carbon nanotubes [11] and graphene-based fillers [10, 11, 16, 17, 20]. Although theories including the Guth-Gold theory [2] and the jamming theory [3, 19] were employed to analyse this observation [2-4, 10-12, 14, 20], there is no specific equation with well-defined parameters, able to describe the reinforcing mechanisms of graphene-reinforced elastomer nanocomposites. Hence, a theoretical analysis based on our recently developed shear-lag/rule-of-mixtures theory, where the parameters are well-defined [21], will be carried out in this work and discussed thoroughly in the next sections.

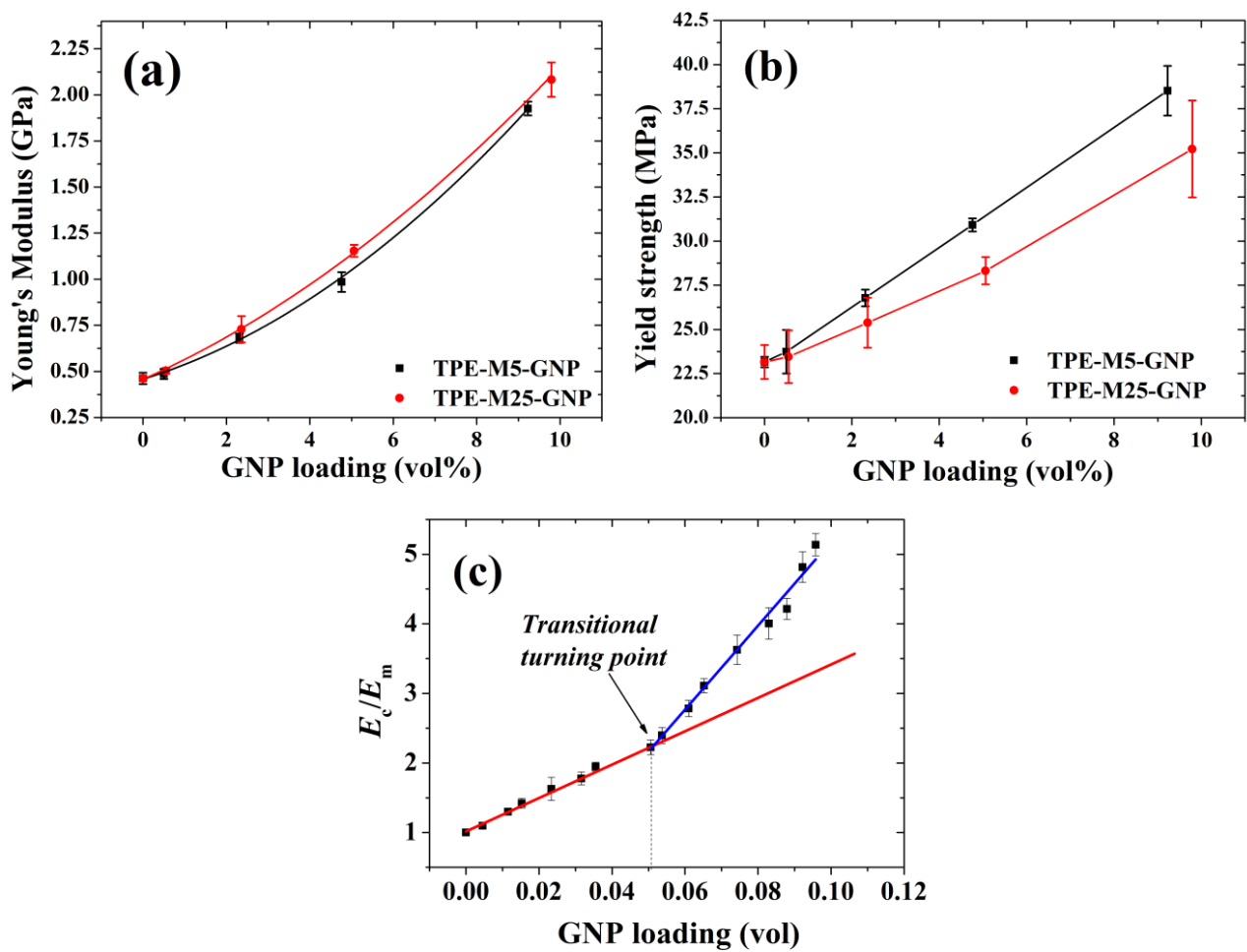

**Figure 2.** (a) Young's modulus and (b) yield strength against volume fractions of the filler (the lines in both (a) and (b) are just a guide to the eye); (c) Young's modulus of the M25-GNP reinforced samples with various filler loadings showing higher reinforcing efficiency at loadings higher than 5 vol%.

*3.4 Raman 2D band shift*

Raman spectroscopy is a powerful technique that is able to investigate the interfacial stress transfer from a polymer matrix to graphene-related materials [8, 13, 18, 21, 24-27]. The specimen with the highest GNP loading (20 wt%) were strained using a bending rig *in situ* under a Raman spectrometer and the characteristic shifts of the Raman bands were recorded with increasing composite strain from 0 to ~2.7%. The corresponding results are shown in Figure 3.

At a strain range from 0 to ~ 1.2 %, the Raman 2D band shifts to lower wavenumbers and the downshift can be fitted linearly in order to obtain the stress transfer efficiency from the matrix to the nanoplatelets [8]. The slope values presented in this work, reveal that the interfacial stress transfer from the matrix to the filler is more efficient compared to a number of softer elastomeric matrices reinforced by GNPs [13, 14, 21], but less efficient compared to stiffer matrices including PP, PMMA and epoxy resins [21, 24-26]. This is in accordance with our recent study [21] where we identified clearly that the filler modulus of graphene in polymer nanocomposites increases almost linearly with increasing matrix modulus.

When the applied load on the specimen was increased from ~1.2% up to ~2.7% strain, the 2D band shift increased and decreased irregularly with increasing strain, indicating a certain degree of relaxation in the specimen [8]. In thermoplastic elastomers, stress relaxation can lead to irreversible disentanglement of the physical crosslinks that are able to support stress for a short time [1].

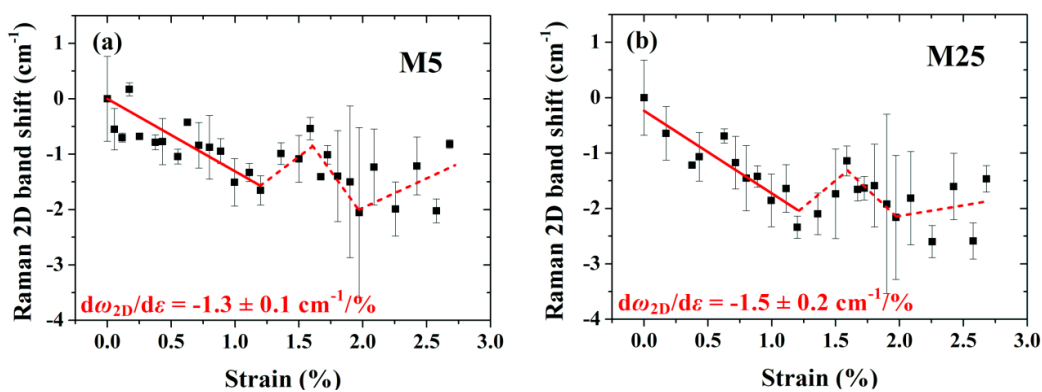

**Figure 3.** Raman 2D band shift against the composites strain of (a) M5 and (b) M25 reinforced TPE composite samples at 20 wt% loading of the filler. The solid lines in both (a) and (b) represent the linear fit of the downshift of the 2D Raman band, for strain up to ~1.2%, while the dashed irregular lines for strain higher than 1.2% are an indication of stress relaxation and disentanglement of the physical crosslinks in the TPE matrix.

The filler modulus can be calculated using the slope of 2D Raman band shift by the equation: $E_R = -\frac{d\omega_{2D}}{d\varepsilon}\frac{1050}{60}$ GPa [8]. The filler modulus results obtained by Raman measurements ($E_R$) and tensile testing ($E_f$) for the samples at the filler content of 20 wt% are listed in Table 1, where $E_f$ was calculated using the simple rule-of-mixtures [1] on the single data point at 20 wt% filler content: $E_c = E_f V_f + E_m(1-V_f)$, where $E_f$ and $E_m$ are the modulus of the filler and the matrix, respectively and $V_f$ is the volume fraction of the filler. It can be seen that the filler modulus measured by the Raman band shift ($E_R$) is similar to the one obtained from tensile tests ($E_f$) in the cases of both M5 and M25 reinforced TPE, with $E_R$ being somewhat higher than $E_f$. The filler modulus determined by tensile testing is based on the overall deformation of the composites. However, during the Raman measurements, the laser spot (in the order of 1-2 µm) generally probes on the centre of the axially-aligned flakes, while the flakes measured are ~5 or ~25 µm. Therefore, the stress at the laser-focused point of the nanoplatelet is higher than the average stress along the flakes, leading to higher filler modulus acquired by Raman measurements ($E_R$) than the filler modulus determined by tensile testing ($E_f$) [21].

**Table 1.** Raman 2D Band shifts and corresponding Raman modulus values along with theoretical Raman modulus values calculated by $E_R=-\frac{d\omega_{2D}}{d\varepsilon}\frac{1050}{60}$ GPa for both M5 and M25 reinforced TPE composites at the highest filler contents.

|  | M5 | M25 |
|---|---|---|
| Band shift (cm$^{-1}$/% strain) | -1.3 ± 0.1 | -1.5 ± 0.2 |
| $E_R$ (GPa) | 22.9 ± 1.8 | 26.3 ± 3.5 |
| $E_f$ (GPa) | 16.6 ± 0.6 | 18.1 ± 1.4 |

## 4. Discussion

### 4.1 Theoretical analysis using micromechanics

The rule-of-mixtures is a well-accepted theory in the field of polymer composites, used to analyze the mechanics of reinforcement [1]. The Young's modulus of the composites is given by:

$$E_c = E_f V_f + E_m (1-V_f) \quad (1)$$

Based on our recent work [21], when elastomers are reinforced by graphene, the $E_f$ is dependent upon the modulus of the matrix, the orientation factor ($\eta_o$) and the aspect ratio (*s*) of the filler [21] as shown in eq. (2):

$$E_f = \eta_o \frac{s^2}{12} \frac{t}{T} \frac{1}{(1+v)} E_m \quad (2)$$

where *s* is the aspect ratio of the filler; *v* is the Poisson's ratio of the elastomer. Moreover, *t* is the thickness of the flake and *T* is the thickness of a layer of the matrix surrounding the flake. If we substitute $E_f$ from equation (2), then equation (1) can be rewritten and rearranged into:

$$E_c/E_m = 1 + (\eta_o \frac{s^2}{12} \frac{t}{T} \frac{1}{(1+v)} - 1)V_f \quad (3)$$

The key to address the issue of the reinforcing mechanism in graphene-reinforced elastomers is to analyze the parameter $t/T$ in equation (3). If we assume a simple model of a single nanoplatelet inserted in a continuous elastomer matrix, then due to the fact that the macromolecular chains of elastomers are very flexible, the matrix unit located at an infinite distance from the flakes cannot be affected [5]. Therefore, $T$ cannot be infinitely large for an elastomer matrix. More specifically, the value of $T$ is dependent upon the shear modulus of the elastomeric matrix, which relates to the flexibility of the macromolecular chain segment. In this context, we assume a shear-lag unit for a single flake in the elastomeric matrix as shown in Figure 4 that can be affected by the flake through shear when an external stress is applied. Any part of the matrix located further than $T$ from the flake cannot be affected.

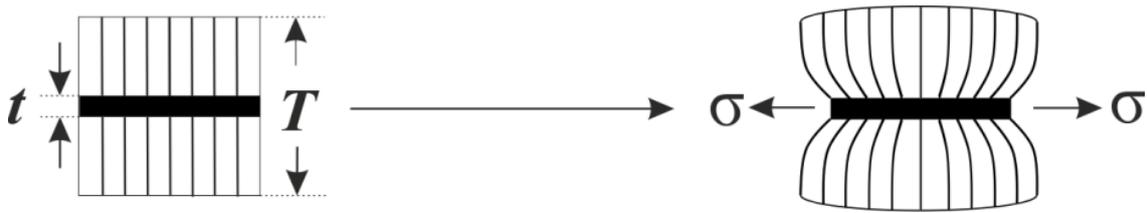

**Figure 4.** (left) Shear-lag stress transfer unit for an individual flake within the elastomer matrix: t is the thickness of the flake and the T is the thickness of the matrix surrounding the flake, which the flake can affect when an external stress is applied; (right) the deformation of the flake and the matrix polymer after application of strain.

The shear-lag unit model illustrates the reinforcement from individual flakes through stress transfer in elastomer nanocomposites. We can now apply the model of multiple shear-lag units into bulk elastomeric nanocomposites. The distribution and behavior of the shear-lag units in elastomer nanocomposites from low to high filler loadings, when an external stress is applied, is demonstrated in Figure 5. In this case, it is assumed that all the nanoplatelets are dispersed homogeneously. When the loading of the filler is low (Figures 5a and b), the flakes are located far enough from each other, making the interaction between shear-lag units impossible. The SEM images of the samples filled with low GNP content, next to the shear-lag unit model, also point towards this direction. In this case, the modulus increase can be attributed to the enhancement from individual flakes, through stress transfer. Hence, the value of $t/T$ is constant and consequently equation (3) shows a linear relationship between the normalized modulus and the volume fraction of the filler. With the increase of $V_f$, the distance between adjacent shear-lag units becomes closer and they finally coincide with each other. At this critical point, the $t/T$ ratio has the same value with the $V_f$ (Figure 5c). We define this point as *the percolation threshold volume fraction* of the filler, $V_p$. It can be understood with the aid of both the model and the SEM Figures 5(a-b) that when the volume fraction of the filler is below $V_p$, the constant parameter, $t/T$, can be approximated to be equal to $V_p$. When the filler content increases from the percolation threshold volume fraction of the filler ($V_p$), to higher filler contents (Figure 5c-d), the thickness of the matrix surrounding the flake ($T$) is geometrically reduced due to the smaller distances between the flakes and the parameter $t/T$ in equation (3) can be eventually substituted by $V_f$. The normalized modulus is then given by:

$$E_c/E_m = 1 - V_f + \eta_o \frac{s^2}{12} \frac{1}{(1+v)} V_f^2 \qquad (4)$$

Assuming the Poisson's ratio of the elastomer is 0.5 [5], equation (3) for filler contents below the percolation threshold takes the form;

$$E_c/E_m = 1 + (0.056 \eta_o s_{eff}^2 \frac{t}{T} - 1) V_f \qquad \text{for} \qquad V_f < V_p \qquad (t/T = V_p) \qquad (5)$$

and equation (4) for filler contents above the percolation threshold can be rewritten as;

$$E_c/E_m = 1 - V_f + 0.056 \eta_o s_{eff}^2 V_f^2 \qquad \text{for} \qquad V_f \geq V_p \qquad (6)$$

where $s_{eff}$ is the effective aspect ratio of the GNPs in the bulk nanocomposites. The schematic diagrams in Figure 5 were drawn for the case of perfect orientation of the filler, in order to simplify the illustration. If the flakes are not perfectly aligned along the direction of the external force, as they are in this case, then the orientation factor $\eta_o$ ($8/15 \leq \eta_o < 1$) which can be obtained experimentally [27], should be taken into account.

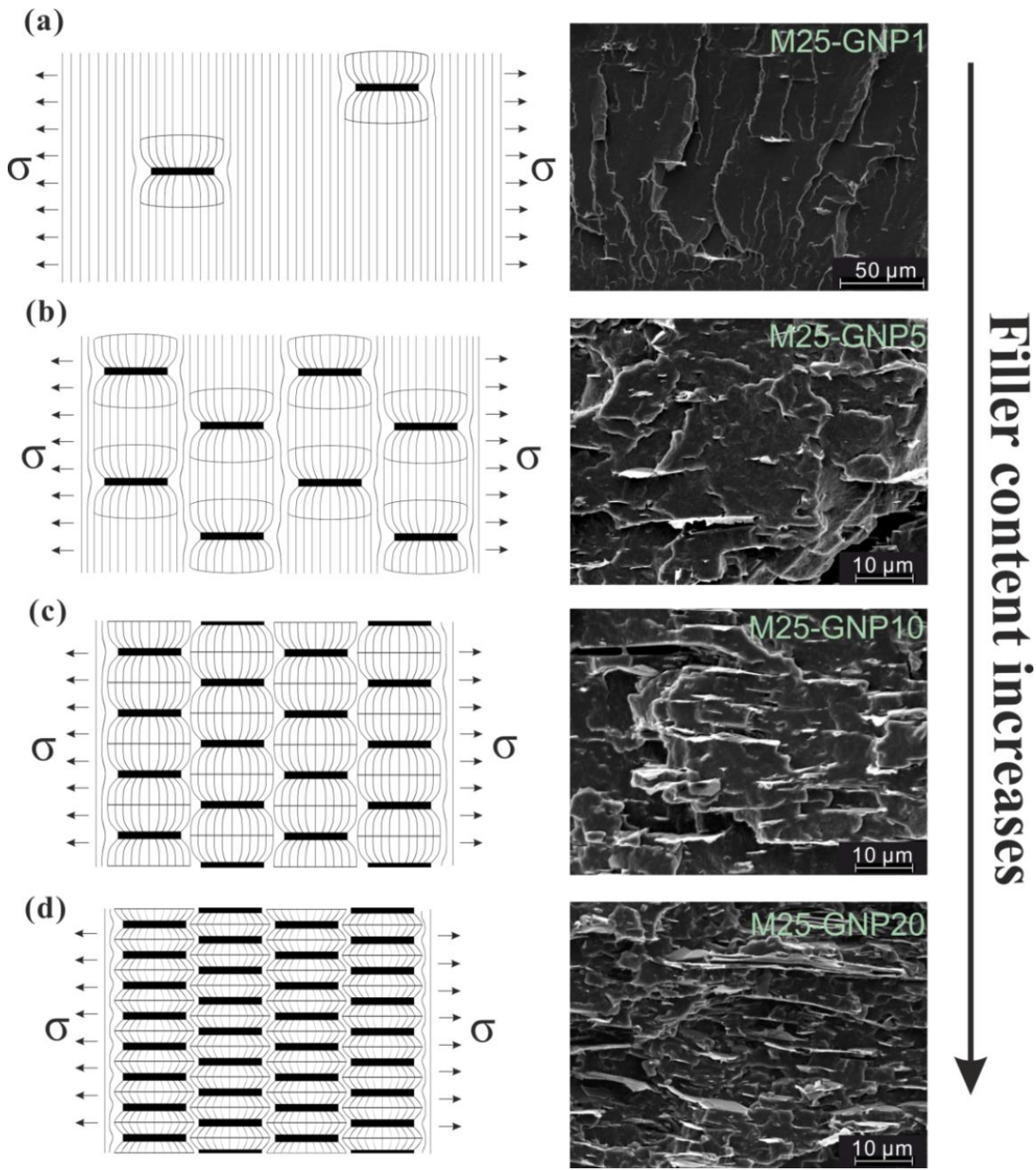

**Figure 5.** Schematic illustration of samples reinforced with different loadings of GNPs (increase from a to d) under external stress, demonstrating the dependence of stress transfer efficiency upon the filler loading in an elastomer matrix, based on shear-lag theory: (a) low filler loading; (b) high filler loading than (a) but below the percolation threshold (both a and b show the reinforcement

from individual flakes); (c) filler loading at percolation threshold and (d) filler loading above the percolation threshold showing the formation of the filler network and the enhanced reinforcing efficiency of the flakes due to the reduced distance between individual flakes. The illustration of the distance between the flakes is also suggested by SEM images as shown.

The proposed equations (5) and (6) can describe the reinforcement from GNPs in elastomer-based nanocomposites, while being able to explain the accelerated stiffening phenomenon with increasing filler content. From low to high filler contents, the reinforcing mechanism of the GNPs in graphene-reinforced elastomer composites can be divided into 3 individual stages:

*Stage I*: When the filler loading is low and below the percolation threshold volume fraction ($V_f < V_p$), it is considered that the mechanical improvements are dependent upon the performance of the individual flakes, where the parameter $t/T$ takes the constant value of $V_p$ and consequently the normalized modulus shows a linear relationship with $V_f$, represented by equation (5).

*Stage II*: With the increase of filler fraction to reach and overcome the percolation threshold ($V_f \geqslant V_p$), the average distance between nanoplatelets is small enough to enable mechanical reinforcement from both individual fillers and the simultaneous contributions by pairs of fillers. This effect can be quantitatively expressed by a quadratic relationship between $E_c/E_m$ and $V_f$, as shown in equation (6), while being also able to explain the accelerated stiffening phenomenon in elastomer composites with increasing filler contents [2].

*Stage III*: When the filler loading is high enough, a number of agglomerates are formed in the nanocomposite and as a result, the reinforcing efficiency is reduced [14, 24, 25]. The influence of agglomeration can be quantitatively realized by the decreased values of the effective aspect ratio, obtained through the fitting of the experimentally obtained modulus data, using equation (6).

*4.2 Application of the proposed theory*

The results of the mechanical properties of the nanocomposites were presented in *Section 3.3* and the fitting of the experimental data using the newly-derived equations (5) and (6), was carried out as shown in Figure 6 for the case of perfect orientation. From the normalized modulus values presented in Figure 2c, it can be clearly observed that the slopes of the experimental data are different before and after the filler content of 5 vol%. On this basis, 5 vol% was established as the percolation threshold volume fraction of the filler ($V_p$). The data below 5 vol%, were fitted by equation (5), where according to the previous discussion, reinforcement depends on stress transfer from individual flakes, while the data above 5 vol% were fitted using equation (6), where the matrix is additionally stiffened.

The data points below 5 vol% were fitted using a linear line, as suggested by equation (5). The slope has the same value as the factor [$0.056\eta_o s_{eff}^2(t/T)-1$], where $t/T$ (=$V_p$) is 0.05. Assuming the orientation of the flakes is perfect ($\eta_o$=1), the effective aspect ratio can be calculated. For the M25 GNPs within the TPE matrix, the value of the effective aspect ratio is 95, as shown in Figure 6. Then, by substituting $s_{eff}$ into equation (6), the fitting of the experimental data is represented by the red curve in Figure 6. It can be seen that equation (6) fits accurately the normalized modulus of the samples filled with ~0.05 – ~0.07 vol. GNPs. The effective aspect ratio (in the order of 100) for M25-GNPs in the bulk composites is relatively low for a 2D material and this can be attributed to the high thickness of the starting material and the presence of looped or folded morphologies amongst the flakes that either pre-exist or are formed as a result of the high shear during the melt mixing process (Figure S4). However, the fitting was not accurate enough for the samples filled with higher GNP loadings, due to the increased number of folded flakes and the unavoidable formation of agglomerates. When adjusting the value of $s_{eff}$ to a slightly lower value ($s_{eff}$=90), the fitting becomes consistent with the data, as shown by the blue-dashed curve in Figure 6. For the case of random orientation ($\eta_o$=0.53), the same fitting procedure can be carried out for the experimental data, as shown in Figure S6. The fitting results display a good consistency with the experimental data, similarly to Figure 6. Moreover, the fitted value of effective aspect ratio is slightly higher, but still in the order of 100.

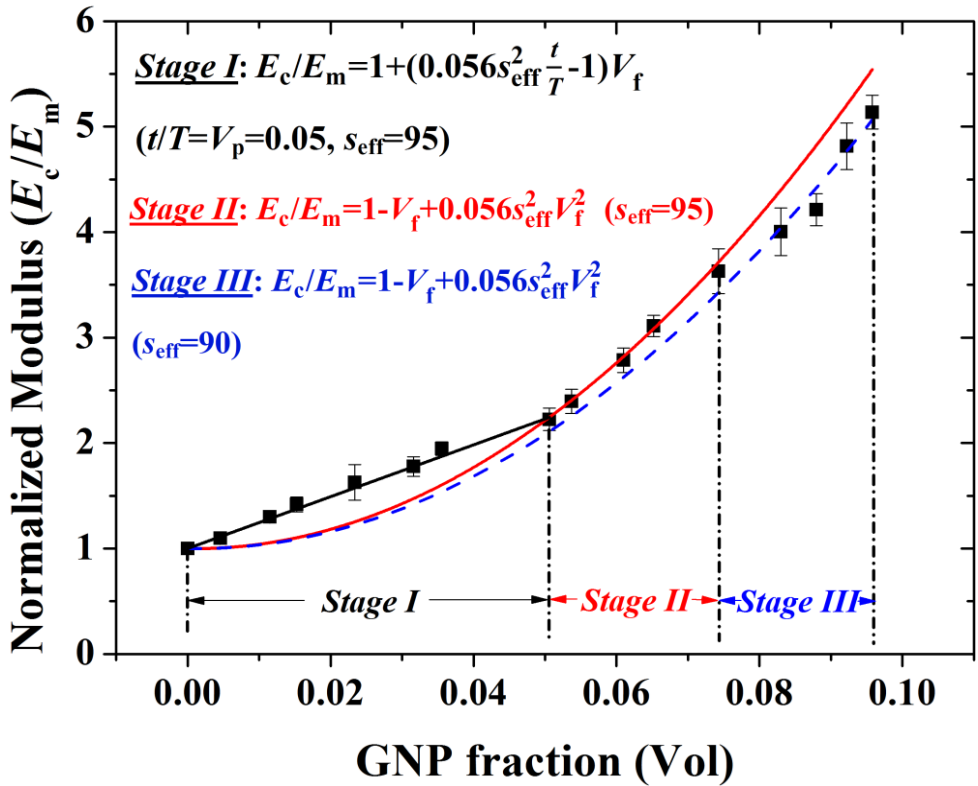

**Figure 6.** Fittings of normalized modulus against volume fraction of the fillers with both equations (5) and (6) for M25-GNP reinforced composite samples showing three stages of the reinforcement, assuming perfect orientation of the flakes. Stage I: reinforcement takes place from individual flakes through stress transfer and the modulus can be fitted with a linear equation (equation 5). Stage II: above the percolation threshold volume fraction, the reinforcement originates from both individual fillers and the simultaneous contributions by pairs of fillers. The effect is described by a quadratic relationship (equation 6). Stage III: higher filler contents lead

to the formation of agglomerates, the modulus can still be fitted with equation 6 however the effective aspect ratio ($s_{eff}$) of the fillers is reduced.

## 5. Conclusions

GNP-reinforced, semi-crystalline thermoplastic elastomers have been successfully prepared by melt mixing. The degree of crystallinity remained constant with the addition of GNPs. SEM images indicate that the dispersion of the GNPs within the matrix is homogeneous, with excellent filler/matrix interfaces in general.

The mechanical properties obtained from tensile testing suggested that the stiffness and yield strength are both significantly enhanced. It was clearly exhibited that after the percolation threshold volume fraction, the modulus of the composites presents a superlinear increase, compared to the linear increase observed at low filler contents. The filler modulus measured by Raman band shifts showed similar, but slightly higher values than the filler modulus determined by tensile testing. Finally, the consistency of the newly-derived equations (5) and (6) with the experimental data has manifested the applicability of the combined shear-lag/rule-of-mixtures theory proposed here, on the reinforcement mechanisms of elastomer/GNP nanocomposites. The three stages of reinforcement from low to high volume fraction of the filler have been clearly identified. Similar analysis methods, considering the mechanical percolation in graphene-based elastomer nanocomposites have been introduced before [10-12, 20], by using either hydrodynamics [2] or the jamming theory [3, 19]. However, an obvious advantage of the theory proposed here, is the accurate interpretation of the percolation of tensile modulus, for the specific case of elastomer nanocomposites reinforced by 2D materials with well-defined parameters.

**Acknowledgements**

This project has received funding from the European Union's Horizon 2020 research and innovation programme under grant agreement No 785219. One of the authors (M. Liu) is grateful to the China Scholarship Council for financial support. All research data supporting this publication are available within this publication.

**References**

[1] Young RJ, Lovell PA. Introduction to polymers: CRC press; 2011.

[2] Guth E. Theory of filler reinforcement. Journal of Applied Physics. 1945;16(1):20-5.

[3] Liff SM, Kumar N, McKinley GH. High-performance elastomeric nanocomposites via solvent-exchange processing. Nature Materials. 2007;6(1):76.

[4] Ramorino G, Bignotti F, Pandini S, Riccò T. Mechanical reinforcement in natural rubber/organoclay nanocomposites. Composites Science and Technology. 2009;69(7-8):1206-11.

[5] Smallwood HM. Limiting law of the reinforcement of rubber. Journal of Applied Physics. 1944;15(11):758-66.

[6] Novoselov KS, Geim AK, Morozov SV, Jiang D, Zhang Y, Dubonos SV, et al. Electric field effect in atomically thin carbon films. Science. 2004;306(5696):666-9.

[7] Lee C, Wei X, Kysar JW, Hone J. Measurement of the elastic properties and intrinsic strength of monolayer graphene. Science. 2008;321(5887):385-8.

[8] Gong L, Kinloch IA, Young RJ, Riaz I, Jalil R, Novoselov KS. Interfacial stress transfer in a graphene monolayer nanocomposite. Advanced Materials. 2010;22(24):2694-7.

[9] Papageorgiou DG, Kinloch IA, Young RJ. Graphene/elastomer nanocomposites. Carbon. 2015;95:460-84.

[10] Araby S, Zaman I, Meng Q, Kawashima N, Michelmore A, Kuan H-C, et al. Melt compounding with graphene to develop functional, high-performance elastomers. Nanotechnology. 2013;24(16):165601.

[11] Das A, Kasaliwal GR, Jurk R, Boldt R, Fischer D, Stöckelhuber KW, et al. Rubber composites based on graphene nanoplatelets, expanded graphite, carbon nanotubes and their combination: a comparative study. Composites Science and Technology. 2012;72(16):1961-7.

[12] Fernández-d'Arlas B, Corcuera MA, Eceiza A. Comparison between exfoliated graphite, graphene oxide and multiwalled carbon nanotubes as reinforcing agents of a polyurethane elastomer. Journal of Thermoplastic Composite Materials. 2015;28(5):705-16.

[13] Li S, Li Z, Burnett TL, Slater TJ, Hashimoto T, Young RJ. Nanocomposites of graphene nanoplatelets in natural rubber: microstructure and mechanisms of reinforcement. Journal of Materials Science. 2017;52(16):9558-72.

[14] Liu M, Papageorgiou DG, Li S, Lin K, Kinloch IA, Young RJ. Micromechanics of reinforcement of a graphene-based thermoplastic elastomer nanocomposite. Composites Part A: Applied Science and Manufacturing. 2018:84-92.

[15] Potts JR, Shankar O, Du L, Ruoff RS. Processing–morphology–property relationships and composite theory analysis of reduced graphene oxide/natural rubber nanocomposites. Macromolecules. 2012;45(15):6045-55.

[16] Wu C, Huang X, Wang G, Wu X, Yang K, Li S, et al. Hyperbranched-polymer functionalization of graphene sheets for enhanced mechanical and dielectric properties of polyurethane composites. Journal of Materials Chemistry. 2012;22(14):7010-9.

[17] Yang L, Phua SL, Toh CL, Zhang L, Ling H, Chang M, et al. Polydopamine-coated graphene as multifunctional nanofillers in polyurethane. RSC Advances. 2013;3(18):6377-85.

[18] Papageorgiou DG, Kinloch IA, Young RJ. Mechanical properties of graphene and graphene-based nanocomposites. Progress in Materials Science. 2017;90:75-127.

[19] Trappe V, Prasad V, Cipelletti L, Segre P, Weitz DA. Jamming phase diagram for attractive particles. Nature. 2001;411(6839):772.

[20] Nawaz K, Khan U, Ul-Haq N, May P, O'Neill A, Coleman JN. Observation of mechanical percolation in functionalized graphene oxide/elastomer composites. Carbon. 2012;50(12):4489-94.

[21] Young RJ, Liu M, Kinloch IA, Li S, Zhao X, Vallés C, et al. The mechanics of reinforcement of polymers by graphene nanoplatelets. Composites Science and Technology. 2018;154:110-6.

[22] Salmoria GV, Paggi RA, Lago A, Beal VE. Microstructural and mechanical characterization of PA12/MWCNTs nanocomposite manufactured by selective laser sintering. Polymer Testing. 2011;30(6):611-5.

[23] Kuo AC. Polymer data handbook. Polymer Data Handbook. 1999.

[24] Ahmad SR, Xue C, Young RJ. The mechanisms of reinforcement of polypropylene by graphene nanoplatelets. Materials Science and Engineering: B. 2017;216:2-9.

[25] Papageorgiou DG, Kinloch IA, Young RJ. Hybrid multifunctional graphene/glass-fibre polypropylene composites. Composites Science and Technology. 2016;137:44-51.

[26] Vallés C, Kinloch IA, Young RJ, Wilson NR, Rourke JP. Graphene oxide and base-washed graphene oxide as reinforcements in PMMA nanocomposites. Composites Science and Technology. 2013;88:158-64.

[27] Li Z, Young RJ, Wilson NR, Kinloch IA, Vallés C, Li Z. Effect of the orientation of graphene-based nanoplatelets upon the Young's modulus of nanocomposites. Composites Science and Technology. 2016;123:125-33.

**SUPPORTING INFORMATION**

*Table S1. Mass fractions of GNPs determined by TGA for each sample and calculated volume fractions.*

| Materials | Mass fraction (%) | GNP volume fraction (%) |
|---|---|---|
| TPE | 0 | 0 |
| TPE-M5-GNP1 | 1.10 ± 0.30 | 0.50 ± 0.14 |
| TPE-M5-GNP5 | 4.95 ± 0.14 | 2.31 ± 0.06 |
| TPE-M5-GNP10 | 9.90 ± 0.14 | 4.76 ± 0.09 |
| TPE-M5-GNP20 | 18.28 ± 0.20 | 9.23 ± 0.18 |
| TPE-M25-GNP1 | 1.20 ± 0.20 | 0.55 ± 0.09 |
| TPE-M25-GNP5 | 5.05 ± 0.03 | 2.36 ± 0.01 |
| TPE-M25-GNP10 | 10.51 ± 0.75 | 5.06 ± 0.34 |
| TPE-M25-GNP20 | 19.28 ± 0.82 | 9.79 ± 0.37 |

The volume fraction was calculated by Equation S1. The volume fraction is given by:

$$V_f = \frac{w_f \rho_m}{w_f \rho_m + (1-w_f)\rho_f} \tag{S1}$$

where $w_f$ is the mass fraction of the filler, $\rho_m$ (=1.01 g/cm$^3$) and $\rho_f$ (=2.2 g/cm$^3$) are the densities of the matrix and the filler, respectively.

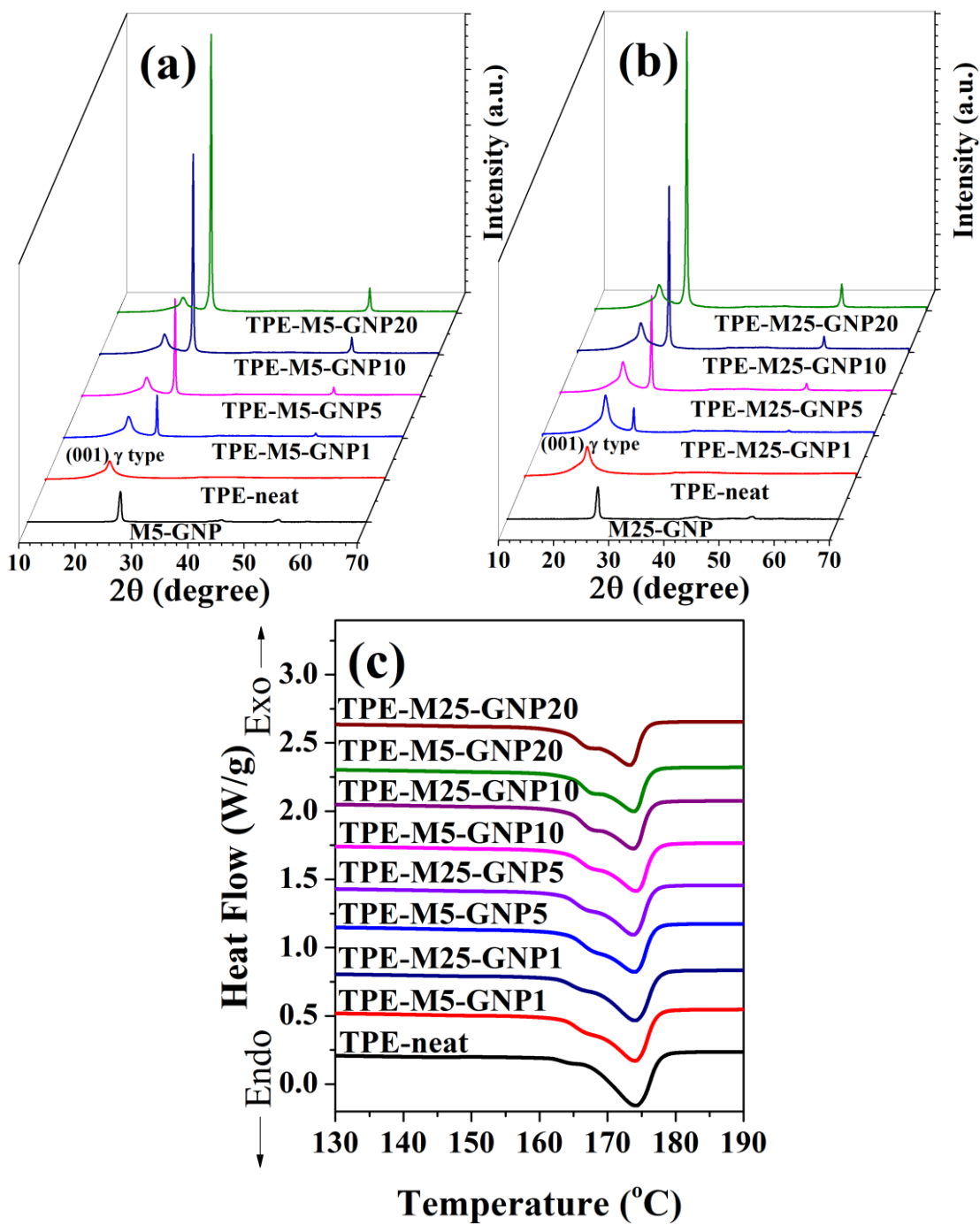

*Figure S1. XRD patterns of (a) M5 and (b) M25 GNP-reinforced TPE and (c) DSC curves from 120 °C to 190 °C of GNP-reinforced TPE samples.*

*Table S2. Degree of crystallinity of the neat polymer and composite samples.*

| Materials | Crystallinity from XRD (%) | Crystallinity from DSC (%) |
|---|---|---|
| TPE | 46.2 ± 0.5 | 47.3 ± 0.7 |
| TPE-M5-GNP1 | 47.4 ± 1.2 | 47.0 ± 1.3 |
| TPE -M5-GNP5 | 47.3 ± 0.8 | 46.9 ± 0.6 |
| TPE -M5-GNP10 | 46.3 ± 1.2 | 47.3 ± 0.6 |
| TPE -M5-GNP20 | 45.4 ± 1.5 | 47.3 ± 0.4 |
| TPE -M25-GNP1 | 48.5 ± 1.2 | 47.2 ± 1.1 |
| TPE -M25-GNP5 | 47.5 ± 0.8 | 47.2 ± 0.5 |
| TPE -M25-GNP10 | 46.6 ± 1.3 | 47.7 ± 0.3 |
| TPE -M25-GNP20 | 45.6 ± 1.4 | 44.7 ± 2.3 |

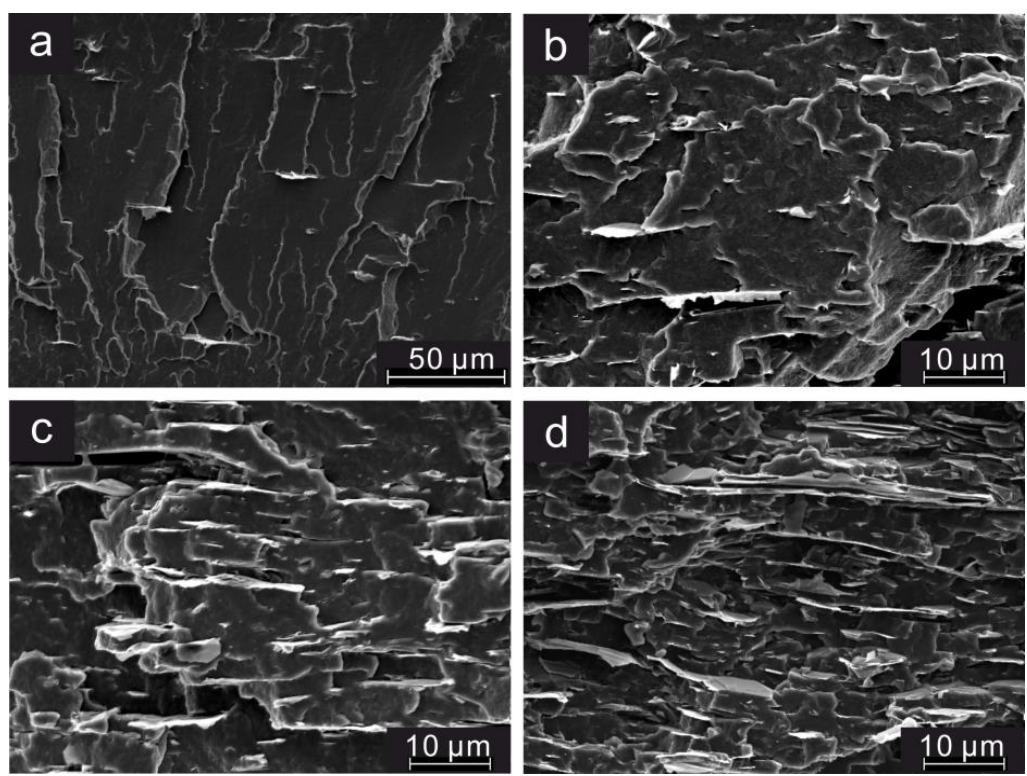

*Figure S2. SEM images of the composites (a) TPE-M25-GNP1, (b) TPE-M25-GNP5, (c) TPE-M25-GNP10, (d) TPE-M25-GNP20.*

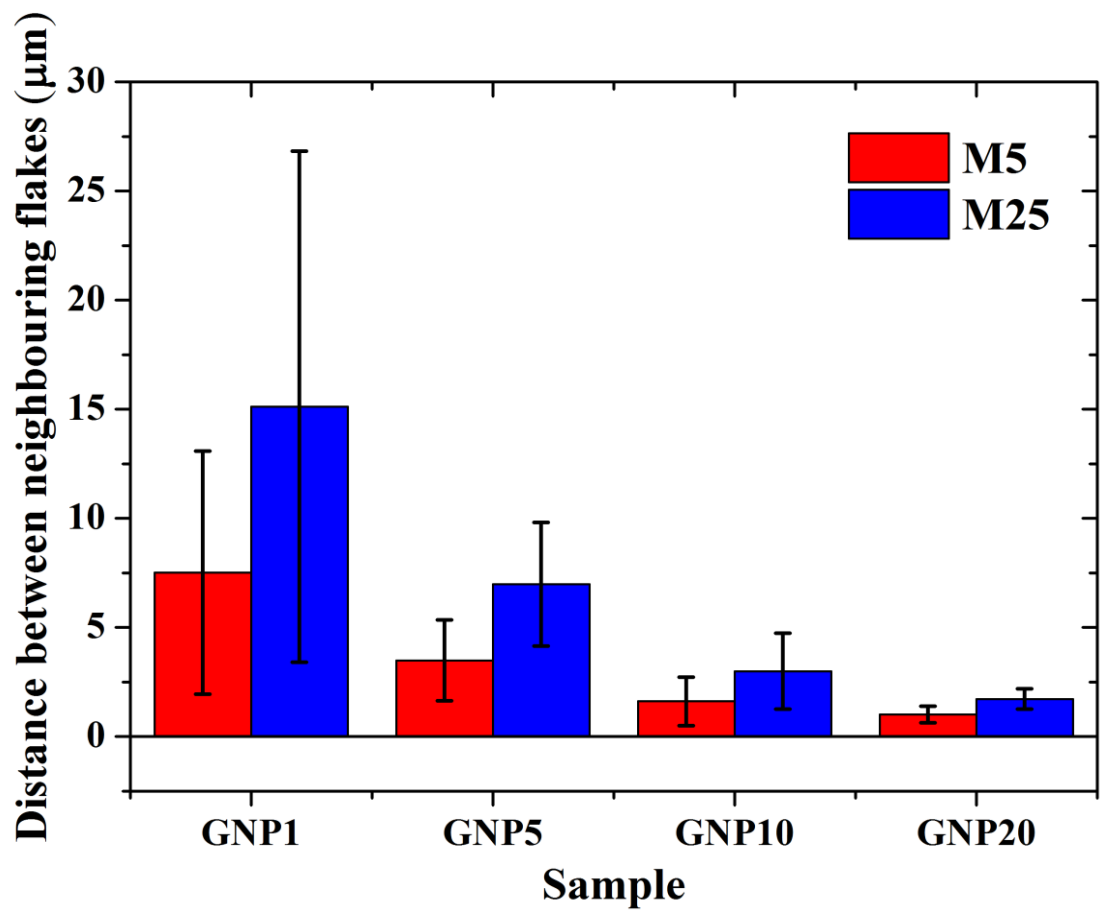

*Figure S3. Measured distance between flakes in surface normal direction based on SEM images for more than 100 flakes each sample.*

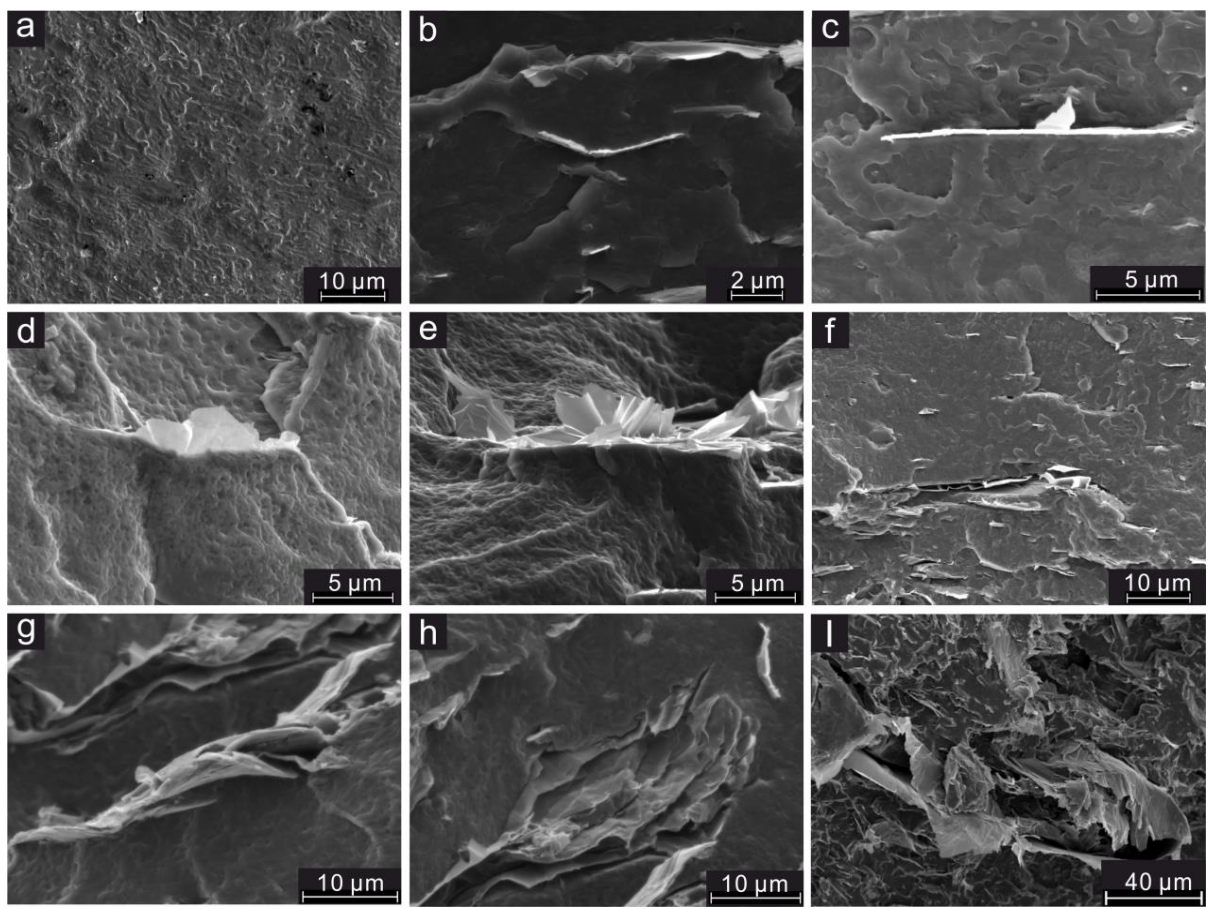

*Figure S4. SEM images of (a) Neat TPE; (b,c) M5-GNP reinforced composites and (d,e) M25-GNP reinforced composites, respectively showing typical flake/matrix interfaces; (f,g) M5-GNP5 and M25-GNP5 samples showing stacking of the flakes; (h,i) M5-GNP5 and M25-GNP10 samples showing agglomerates.*

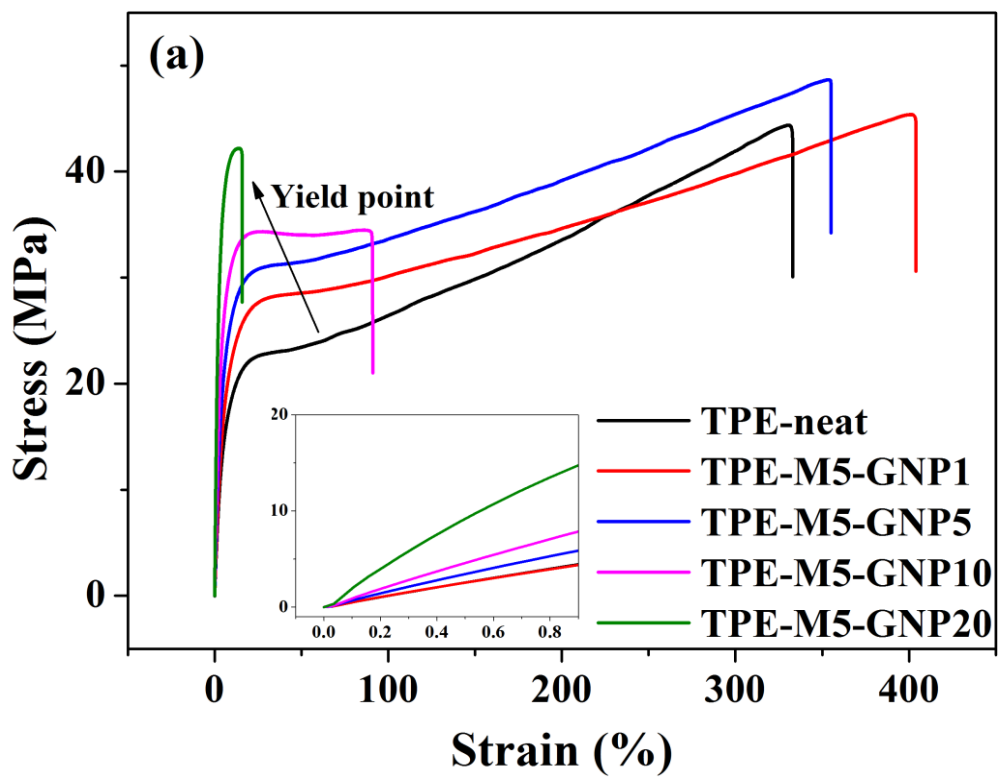

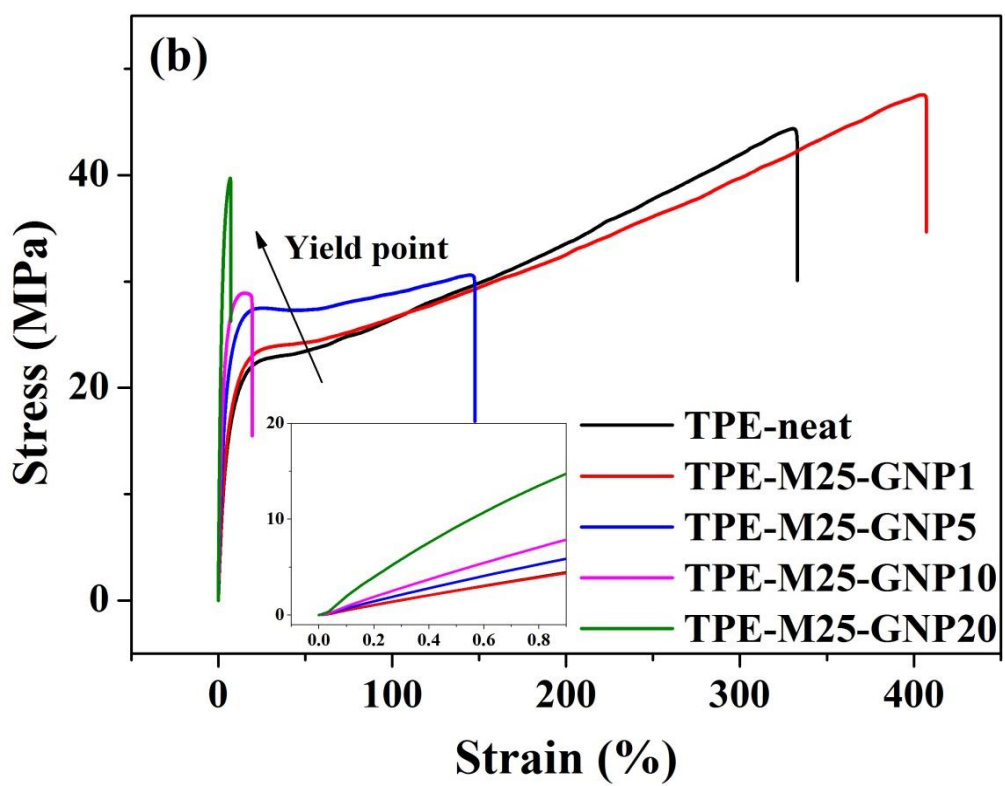

*Figure S5. Typical stress-strain curves for (a) M5 GNP-reinforced and (b) M25 GNP-reinforced elastomer nanocomposites; the insets are the stress-strain curves at low strain.*

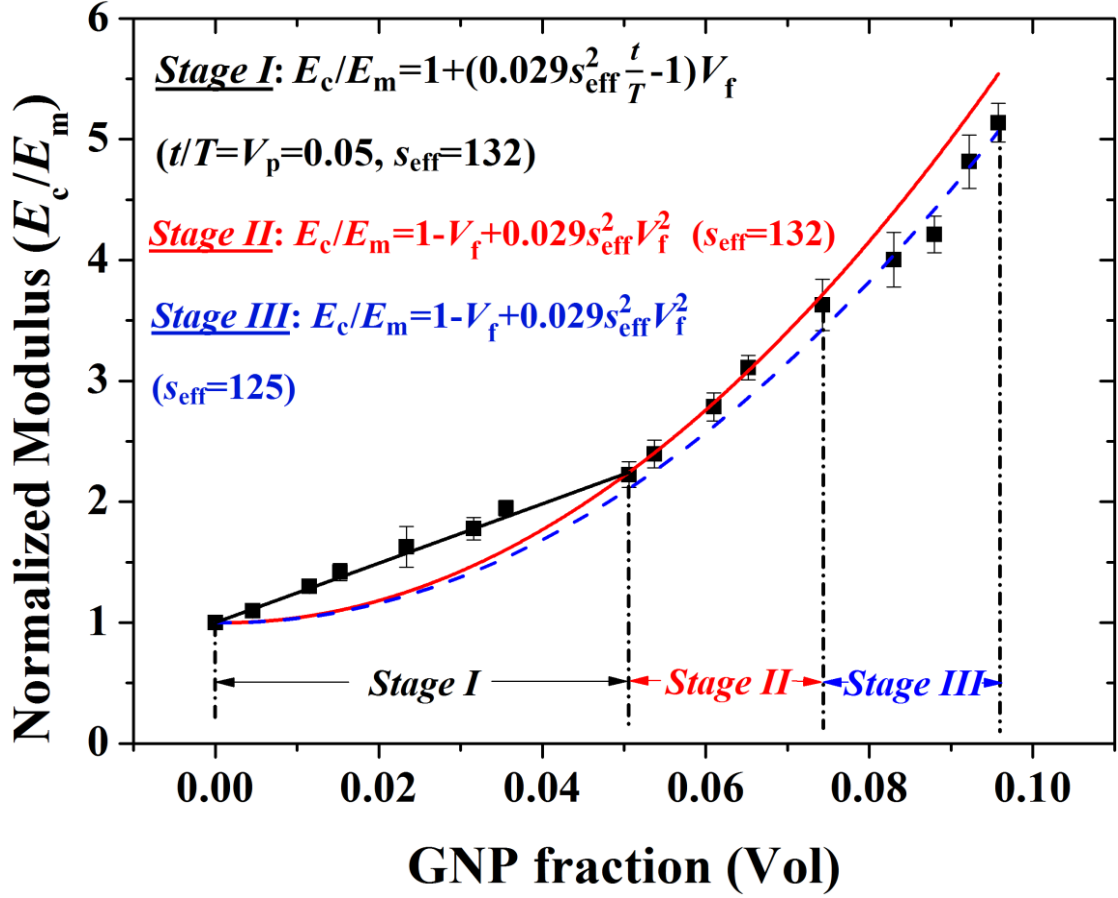

*Figure S6. Fittings of normalized modulus against volume fraction of the fillers with both eq. (5) and (6) for M25-GNP reinforced composite samples showing three stages of the reinforcement, assuming random orientation of the flakes.*